\documentclass[11pt]{article}

\RequirePackage{fullpage}

\usepackage{graphicx}

\title{When can relative risks provide causal estimates?}

\author{A.J. Webster}

\begin{document}

\maketitle

\begin{abstract}
It is emphasised that for epidemiological studies where  
disease incidence is rare, results from conventional proportional 
hazards models can often correctly estimate causal associations. 
The well-known ``backdoor criteria'' from causal-inference is applied to the 
common epidemiological study of rare diseases with a proportional hazards 
model, providing an example of 
when and how estimates from conventional 
proportional hazards 
studies can be used. 
A similar study with the ``frontdoor criteria'', that allows studies with 
unmeasured confounders, finds similar results to conventional mediation 
analysis with measured confounders. 
Reasons for this are discussed. 
\end{abstract}

\section*{Causal inferences}

Whereas statistics is the science of finding and describing patterns in data, 
epidemiology is the science of using statistics to make correct inferences.
Although epidemiologists are careful to describe their results in terms of 
``associations'', the purpose of epidemiology is to detect and quantify 
causal associations, e.g. between lifestyles and health. 
Recently the science of causal inference \cite{Pearl, PearlIntro, VanDerWeele}, 
has developed to allow estimates of causal associations to be made,  
even when the data are not from randomised control trials (RCTs).
By exploring how causal estimates made using the ``backdoor criteria'' 
and the ``do'' calculus \cite{Pearl, PearlIntro}, relate to conventional  epidemiological estimates of relative risks using proportional hazards \cite{Collett2014},  
this short note observes that for many epidemiological studies, 
many conventionally estimated associations \cite{Collett2014, modEpi} 
will correctly estimate causal associations. 
It is also shown how proportional hazards estimates can be applied to 
situations that satisfy the ``frontdoor'' criteria, and how the results can 
coincide with mediation analyses \cite{VanDerWeele}. 

\section*{Causal estimates and relative risks - the ``backdoor criteria''}

\begin{figure}[ht]
	\centering
	\includegraphics[width=1.\linewidth,angle=0]{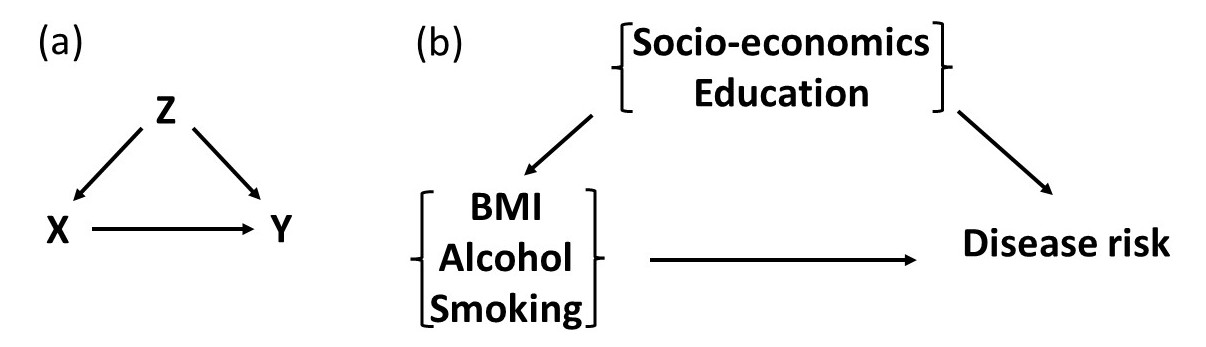}
	\caption[]{
		Consider the influence of one or more exposures 
		$X$, on diseases $Y$, with confounding variables $Z$ that 
		satisfy the ``backdoor criteria'' \cite{Pearl, PearlIntro} (figure a).
		For example, $X$ might include BMI, alcohol, and smoking, with 
		confounders $Z$ of socio-economic status and education (figure b).    
	}\label{XYZ}
\end{figure}

Consider the example of how disease risk may be influenced by the 
established common risk factors of 
education, socio-economic status, BMI, alcohol, and smoking (figure \ref{XYZ}).
It seems likely that for many diseases, education and socio-economic status could 
influence disease risk through the modifiable risk factors of BMI, alcohol, and smoking, in addition to any direct risk. 
In those circumstances education and socio-economic status are confounders 
(denoted with a vector $Z$), that influence both disease risk and the 
values of BMI, alcohol, and smoking (denoted with a vector $X$). 
For this causal model illustrated in figure \ref{XYZ}, it is possible to 
estimate the consequences of setting BMI, alcohol, and smoking to a 
specific value $X=x$, corresponding to $\mbox{do}(X=x)$ using 
the ``do" notation of Pearl \cite{Pearl, PearlIntro}. 
The situation corresponds to the well-known situation described by 
the ``adjustment'' formula, that states, 
\begin{equation}\label{bdoor}
P(Y=y|\mbox{do}(X=x)) = 
\sum_z P(Y=y|X=x, Z=z)P(Z=z)
\end{equation}
where for continuous variables, the sums are treated as integrals. 
The formula is constructed to account for the confounding influence of $Z$ on  
both $X$ and disease risk, and differs from that given by conventional probability 
theory, that would have $P(Z=z|X=x)$ instead of $P(Z=z)$. 
Take $Y=1$ to denote presence of disease, and $Y=0$ its absence, so that, 
\begin{equation}
P(Y=1,T<t | X=x,Z=z)=F(t,x,z) 
\end{equation}
where $F(t,x,z)$ is the distribution function (with covariates $x,z$), so that, 
\begin{equation}
\begin{array}{ll}
F(t,x,z) 
&= 1 - S(t,x,z)
\\
&= 1- \exp(-H(t,x,z)) 
\\
&\simeq H(t,x,z) 
\\
&=e^{\eta_x  + \eta_z } H_0(t) \hspace{.3cm} \mbox{ with } \hspace{.3cm} 
H_0(t) = \int_0^t h_0(s) ds 
\end{array}
\end{equation}
where in going from the 2nd to 3rd line we assume sufficiently rare diseases, as 
is the case for the first incidence of most diseases in 
UK Biobank \cite{UKBiobank}, and in 
going from the 3rd to the 4th lines we assume that the proportional hazards 
assumption \cite{Collett2014} is valid for the disease being studied, 
with $\eta_x$ and $\eta_z$ being the linear predictor 
functions\footnote{For a particular $X=x$, the linear predictor function 
	is sometimes referred to as the ``linear component'', ``risk score'', or ``prognostic index'' \cite{Collett2014}.}
for the (possibly) multivariate 
variables $x$ and $z$.  
For a probability density $f=dF/dt$ and hazard function $h=f/S$, 
$H(t)=\int_0^t h(t)$ is the cumulative hazard, and a proportional hazards 
model assumes that $h(t,x,z)=h_0(t)e^{\eta_x+\eta_z}$. 
Now using Eq. \ref{bdoor}, 
\begin{equation}\label{pydox}
\begin{array}{l}
P(Y=1,T<t | \mbox{do}(X=x) )
\\
= \sum_z P(Y=y, T<t | X=X, Z=z) P(Z=z)
\\
\simeq \sum_z e^{\eta_x  +\eta_z } H_0(t) P(Z=z)
\\
= e^{\eta_x} H_0(t) A_Z
\end{array}
\end{equation}
with $A_Z \equiv \sum_z e^{\eta_z}P(Z=z)$. 
This allows the incidence rates to be calculated for a (possibly hypothetical)  situation where we have intervened in some way to set $X=x$, in terms of a 
baseline hazard function that is estimated in the usual way, using observational 
data in which $Z$ can be correlated with both $X$ and disease risk. 
Note that $P(Z=z)$ and $P(X=x)$ are implicitly the population values 
at the study's start. 
At the baseline values of $x=x_0$ and $z=z_0$, by definition 
$\eta_x(x_0)=0$ and $\eta_z(z_0)=0$, so Eq. \ref{pydox} 
gives $P(Y=1,T<t|X=x_0,Z=z_0)=H_0(t)$. 
Then with the same approximations used to derive Eq. \ref{pydox}, Eq. 
\ref{pydox} can be written in several different ways, for example with,  
\begin{equation}\label{alternateEq}
\begin{array}{l}
P(Y=1,T<t | \mbox{do}(X=x) )
\\
=e^{\eta_x} A_Z H_0(t) 
\\
=e^{\eta_x} A_Z P(Y=1,T<t | X=x_0,Z=z_0 )
\\
=A_Z  P(Y=1,T<t | X, Z=z_0 )
\end{array}
\end{equation}
When education and socio-economic factors are represented by $Z$, then the 
factor $A_Z$ accounts for changes in risk due to both socio-economic factors 
and education, and the influence of setting $X=x$ is calculated through 
the factor $e^{\eta_x}$.  
If we could set $X$ equal to the baseline values $x_0$, the probability 
distribution would be proportional to the baseline hazard function 
$H_0(t)$, amplified/shrunk by the factor $A_Z$. 
If the baseline values corresponded to the lowest disease 
risk, then  $A_Z H_0(t)$ would be the lowest possible disease incidence 
rate that could have been achieved through lifestyle changes. 
Eq. \ref{alternateEq} can also be written as, 
\begin{equation}\label{final1}
\frac{P(Y=1,T<t| \mbox{do}(X=x))}{P(Y=1,T<t| \mbox{do}(X=x_0))}
= e^{\eta_x} 
\end{equation}
This gives the relative risk of disease within time $t$ for a 
population with $X=x$, compared with a population with baseline 
values of $X=x_0$, in terms of estimates from a relative risk 
from observational studies, that have,  
\begin{equation}
e^{\eta_x} 
= \frac{h(t|X=x,Z=z_0)}{h(t|X=x_0,Z=z_0)}
\end{equation}
Using Eq. \ref{pydox} we can calculate the probability density function, 
\begin{equation}
\begin{array}{ll}
P(Y=1,T\in(t,t+\delta t)| \mbox{do}(X=x)) 
&= \delta t
\frac{d}{dt} P(Y=1,T<t| \mbox{do}(X=x))
\\
&\simeq e^{\eta_x} A_Z h_0(t) \delta t 
\\
&= e^{\eta_x} A_Z P(Y=1,T\in(t,t+\delta t)| X=x_0, Z=z_0) 
\end{array}
\end{equation}
This allows a hazard function (usually defined as $h=f/S=f/(1-F)$), 
to be defined as, 
\begin{equation}
 h(t|\mbox{do}(X=x)) \delta t = 
\frac{P(Y=1,T\in(t,t+\delta t)| \mbox{do}(X))}
{1-P(Y=1,T<t| \mbox{do}(X))}
\simeq e^{\eta_x} A_Z  h_0(t) \delta t
\end{equation}
Alternately, we could have argued that most diseases in UK Biobank are sufficiently 
rare that we can approximate $f(t)\simeq h(t)$, and hence that 
$h(t|\mbox{do}(X=x)) \delta t \simeq P(Y=1,T\in(t,t+\delta t)| \mbox{do}(X=x))$
Therefore for combinations of confounders and risk factors that 
satisfy the ``backdoor criteria'' \cite{PearlIntro} 
(such as those in figure \ref{XYZ}),  
and disease incidence that is sufficiently rare  
(which includes most studies of the first incidence 
of a disease in UK Biobank data), 
\begin{equation}\label{RRest}
\frac{h(t|\mbox{do}(X=x))}{h(t|\mbox{do}(X=x_0))} 
	\simeq  e^{\eta_x}
\end{equation}
This indicates that if we are interested in the relative difference in disease rates that would be caused by changing 
from the baseline value $X=x_0$, to $X=x$ (e.g. to reduce risk), then the result is given in terms of the 
usual relative risk, but without terms in $Z$ (corresponding to $Z=z_0$). 
This indicates that causal inferences for $x$ using the relative risk, will be correct in these circumstances.  
These remarks do not apply to any confounders $Z\neq z_0$ that may appear 
in an estimated relative risk, that are set at their 
baseline values $Z=z_0$ to estimate Eqs. \ref{RRest} and \ref{final1}. 
Note that hazard functions do not give a probability in the 
conventional sense, and are not normalised to $1$ for example.  
To ensure the above calculations were done correctly and will 
e.g. normalise to $1$ (within the limits of the approximations made), 
it was essential for the arguments to have 
used probabilities (for which the backdoor theorem applies). 
The remarks apply more generally than to the specific example 
shown in Figure \ref{XYZ}, and to studies other than those involving 
disease or health. 
Similar results will apply whenever $F(t,x,z)$ can be factored as 
$H_0(t)g(x)q(z)$, for some functions $g(x)$ and $q(z)$, as was possible 
here because we consider a proportional hazards model and situations 
whose the incidence is sufficiently rare that we can approximate 
$F(t,x,z) \simeq H(t,x,z)$. 
The above equations can also be used to calculate several related expressions, 
such as the population attributable fraction.
For a clear exposition of when a particular situation will satisfy the 
``backdoor criterion'', please refer to the references \cite{Pearl, PearlIntro}. 

\section*{Unmeasured confounders and mediation - the ``frontdoor criteria''}

\begin{figure}[ht]
 	\centering
 	\vspace*{-0mm}
 	\hspace*{-15.mm}
 	\includegraphics[width=0.7\linewidth,angle=0]{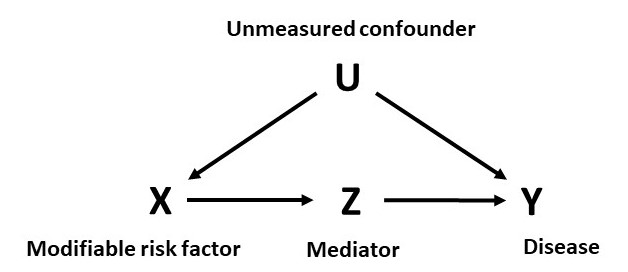}
 	\caption[]{
 		The ``frontdoor criteria'' estimates the causal influence of 
 		an exposure do$(X=x)$, that is mediated by $Z$, in the presence 
 		of unmeasured confounders $U$ that influence both the disease risk 
 		and the exposure $X$. 
 	}\label{medFig}
\end{figure}

Another important result for causal inference, 
is the ``frontdoor criteria'' \cite{Pearl, PearlIntro}. 
A well-known example \cite{PearlIntro} is the assessment of the influence of 
smoking on disease risk in the presence of {\sl unmeasured} confounders 
that influence both smoking use and disease risk, by having an additional 
measurement of tar in peoples' lungs (figure \ref{medFig}). 
Again we consider the adjustment formula for this situation in the limit of 
rare diseases, as above, and consider the simple specific example with 
continuous variables for e.g. average number of cigarettes per day and 
tar content of lungs. 
Although the estimated incidence rates will be found to differ from 
those using proportional hazards models, the causal estimate for the 
smoking and tar example, is the same as we might (with hindsight) have 
anticipated from mediation studies.

For the situation described in figure \ref{medFig}, the ``front door'' adjustment  
formula states \cite{Pearl, PearlIntro},  
\begin{equation}
P(Y=y|\mbox{do}(X=x)) = 
\sum_z \sum_{x'} P(Y=y|Z=z,X=x')P(X=x')P(Z=z|X=x)
\end{equation}
Using this, and proceeding as before, 
\begin{equation}\label{fdEq}
\begin{array}{ll}
P(Y=1,T<t | \mbox{do}(X=x)) 
&= \sum_z \sum_{x'} P(Y=1,T<t|X=x',Z=z) P(Z=z|X=x) P(X=x') 
\\
&\simeq \sum_z \sum_{x'} e^{\eta_{x'}} e^{\eta_z} H_0(t) P(Z=z|X=x)P(X=x')
\\
&= H_0(t)
 \left( \sum_z e^{\eta_z} P(Z=z|X=x) \right) 
 \left(\sum_{x'} e^{\eta_{x'}} P(X=x') \right) 
\end{array}
\end{equation}
Next consider the specific example where 
$\eta_{x'} = \beta_x x'$, 
$\eta_z=\beta_z z$, 
$P(X=x)$ is a normal distribution  $N(\mu_x,\sigma_x^2)$, and 
$P(Z=z|X=x)$ is a normal distribution $N(\alpha x,\sigma_z^2)$, 
where in the latter case $\alpha$ is a constant and the mean of $z$ is 
$\alpha x$. 
Understanding that the sums should be considered as integrals 
when variables are continuous, then we have, 
\begin{equation}
\sum_{x'} e^{\eta_{x'}} P(X=x') = 
	\exp\left(\beta_x \mu_x \right)
	\exp\left( \frac{\sigma_x^2\beta_x^2}{2} \right)
\end{equation}
and, 
\begin{eqnarray}
\sum_{z} e^{\eta_{z}} P(Z=z|X=x) = \exp\left(\beta_z \alpha x \right)
	\exp\left( \frac{\sigma_z^2\beta_z^2}{2} \right)
\end{eqnarray}
giving, 
\begin{equation}
P(Y=1,T<t | \mbox{do}(X=x)) 
\simeq 
H_0(t) \exp(\beta_x\mu_x) 
	\exp\left( \frac{\sigma_x^2\beta_x^2}{2} + \frac{\sigma_z^2\beta_z^2}{2} \right)
	\exp(\beta_z \alpha x)
\end{equation}
The incidence rate at baseline $X=x_0$ is determined by the first three terms, 
and differs from 
a proportional hazard estimate that is adjusted by 
either or both, of $x$ or $z$. 
The first two terms are equivalent to a proportional hazards estimate with 
$x$ at the mean exposure $\mu_x$ and $z$ at the baseline value, 
and the third term quantitatively accounts for the spread in values of $x$ 
and $z$ about their mean values. 
The influence of do$(X=x)$, is seen in the last term $e^{\beta_z\alpha x}$, 
with the change in risk being mediated by $z$ in a very simple and intuitive 
way.
For the situation considered here, 
where there is solely an indirect effect of the exposure through the mediator, 
this estimate is the same as for a mediation analysis with {\sl measured} 
confounding \cite{VanDerWeele}. 
Interestingly, in the equivalent mediation analysis with measured confounding, 
the influence of measured confounding on the 
estimate\footnote{For a solely indirect effect, $\gamma_1=\gamma_3=0$ in 
	Eq. 4.6 on page 101 of 
	\cite{VanDerWeele}, and measured confounding is accounted for through the 
	coefficient $\gamma_4$, that does not subsequently appear in the 
	equations for natural direct and indirect effects.}, 
	does not appear in the resulting expressions for natural direct, 
	and indirect, effects. 
This appears to explain the agreement between estimates with measured, and 
unmeasured confounding - for the model of figure \ref{medFig} in limit of 
rare diseases and a proportional hazards model, the estimate is (apparently) 
unaffected by confounding.
Equation \ref{fdEq} applies to any situation described by figure \ref{medFig}, 
and the example given can be generalised, e.g. to multivariate normal distributions. 

\section*{Conclusions}

The main purpose of this article is to draw attention to the fact that 
many epidemiological studies will, in principle, provide correct estimates 
of causal associations.
This requires a sufficiently good model of the data, and the correct adjustment 
for all confounders. 
The first example that was discussed, was a causal model with risk 
factors $X$ that satisfied the ``backdoor criteria'' \cite{PearlIntro}, 
a situation that is likely to commonly occur. 
This would be important for any meta-analyses of data that are not from 
randomised control trials. 
Often some of the reported estimates will represent the causal influence 
of a potential risk factor and could be included, such as BMI, alcohol, 
and smoking in the first example considered here, but rarely will this be 
true of all variables that are adjusted for. 
Most importantly, where a study has inappropriately adjusted for potential 
confounding variables, then the data cannot be included in the meta-analysis. 
This requires a good causal understanding of how the risk 
factor of interest modifies disease risk, and the potential confounding 
factors that need adjusting for. 
Disagreement between studies may indicate incomplete understanding of the 
underlying causal model, with inappropriate or insufficient adjustment for 
confounding factors. 
In the common situation where uncertainty of the causal processes 
linking exposure $X$ to disease risk remain, then the standard 
methods \cite{modEpi}, 
and cautious reporting of conventional epidemiology must remain. 
The second example considered the ``frontdoor'' criteria, that can allow causal 
estimations in the presence of unmeasured confounders, when the influence of 
$X$ is mediated by a measurable variable $Z$. 
In this case, for the simple example considered 
(with rare disease incidence, and the linear influence of a continuous 
exposure mediated by a normally distributed continuous variable whose 
mean is linearly related to the exposure), 
the causal estimate is the same as you get from a mediation analysis 
with measured confounders. 
With hindsight, this might have been anticipated by observing that 
for this situation (figure \ref{medFig}), 
in the limit of rare diseases studied with a proportional hazards model,  
the confounding terms do not appear in the results of mediation studies 
for natural, and indirect effects \cite{VanDerWeele}. 

\end{document}